# Intelligent Identification of Two-Dimensional Structure by Machine-Learning Optical Microscopy

*Xiaoyang Lin[†,∥,‡,\*], Zhizhong Si[†,‡], Wenzhi Fu[□,‡], Jianlei Yang[□,‡], Side Guo[†], Yuan Cao[†], Jin Zhang[§], Xinhe Wang[†,§], Peng Liu[§], Kaili Jiang[§], Youguang Zhang[†], Weisheng Zhao[†,∥,\*]*

[†] Fert Beijing Research Institute, School of Electrical and Information Engineering & Beijing Advanced Innovation Center for Big Data and Brain Computing (BDBC), Beihang University, Beijing 100191, China

[∥] Beihang-Goertek Joint Microelectronics Institute, Qingdao Research Institute, Beihang University, Qingdao 266000, China

[□] Fert Beijing Research Institute, School of Computer Science and Engineering & Beijing Advanced Innovation Center for Big Data and Brain Computing (BDBC), Beihang University, Beijing 100191, China

[§] State Key Laboratory of Low-Dimensional Quantum Physics, Department of Physics & Tsinghua-Foxconn Nanotechnology Research Center, Collaborative Innovation Center of Quantum Matter, Tsinghua University, Beijing 100084, China

[‡]These authors contributed equally.

\*E-mail: XYLin@buaa.edu.cn (X.Y.L), weisheng.zhao@buaa.edu.cn (W.S.Z)

**Two-dimensional (2D) materials and their heterostructures, with wafer-scale synthesis methods and fascinating properties, have attracted numerous interest and triggered revolutions of corresponding device applications. However, facile methods to realize accurate, intelligent and large-area characterizations of these 2D structures are still highly desired. Here, we report a successful application of machine-learning strategy in the optical identification of 2D structure. The machine-learning optical identification method (MOI method) endows optical microscopy with intelligent insight into the characteristic colour information in the optical photograph. Experimental results indicate that the MOI method enables accurate, intelligent and large-area characterizations of graphene, molybdenum disulphide (MoS$_2$) and their heterostructures, including identifications of the thickness, the existence of impurities, and even the stacking order. Thanks to the convergence of artificial intelligence and nanoscience, this intelligent identification method can certainly promote the fundamental research and wafer-scale device application of 2D structures.**

## Introduction

Two-dimensional (2D) materials have attracted increasing interest, due to their superior properties[1-4]. Heterostructures of 2D materials, which enable great flexibility in both the junction fabrication and property engineering, have further triggered revolutions of corresponding device applications[5-8]. Considering the progress of wafer-scale synthesis method[9-11], the development of an efficient and large-area characterization technique has been a primary obstacle for the fundamental research and commercial-level application of these 2D structures. Among the existed techniques, transmission electron microscopy (TEM) and scanning tunnelling microscopy (STM) enables high spatial-resolution characterization down to atomic scale.[12-16] However, both of them suffer from drawbacks of low-throughput and complicated sample preparation. Atomic force microscopy (AFM) with special design can also enables atomic characterization of 2D materials and even the interface of 2D heterostructures, although still stuck with the surface adsorbates[17,18]. Optical spectroscopy, for example, Raman spectroscopy, can realize accurate characterization of 2D structures[19,20]. However, the spectroscopy method usually enables local characterization within the light spot, resulting in a limited efficiency. Compared to aforementioned techniques, optical microscopy methods, which enable high-speed, large-area, non-destructive and accurate identification of the sample from the as-collected optical photographs (*that is*, the ability of wide-field characterization), has already boosted the controllable synthesis or fabrication, structure-dependent physical property measurement, and device application of 2D structures[21-27].

Optical microscopy method by characterizing bright-field photographs of 2D materials has been applied for the large-area or even wafer-scale characterization of 2D materials[8,9,26-29]. Recently, identification of interlayer coupling in 2D vertical heterojunctions has successfully extended this characterization method to 2D heterostructures, although it relies on a modified optical microscope with the ability of photoluminescence (PL) imaging[30,31]. However, there are still two drawbacks: (1) identifications of 2D heterostructures by optical microscopy are still immature; (2) the optical microscopy method often relies on the experience of the user. Unless intelligent image processing and identification of 2D structures can be realized, these drawbacks can greatly hamper its applications. Adoption of machine-learning strategy in image identification or visual recognition have achieved distinct advantages over humans, which implies great potential of artificial intelligence in image identification of micro and especially nano structures[32-35]. In this sense, integration of machine-learning strategy with optical microscopy technique may realize accurate, intelligent and large-area characterizations of 2D materials and even 2D heterostructures, which can further promote both the fundamental research and commercial-level application.

In this work, we apply the machine-learning strategy in the optical identification of 2D structures, including graphene, molybdenum disulphide ($MoS_2$) and heterostructures of these two materials. The machine-learning optical identification method (MOI method) relies on trainable and automatic analyses of the red, green and blue (RGB) information in the optical photograph of 2D structures by a support vector machine (SVM) algorithm. With this intelligent insight into the characteristic colour information of 2D structure, the MOI method enables accurate and intelligent characterizations of 2D structures, including identifications of the thickness, the existence of impurities, and even the stacking order, which may promote the development of 2D science and technology.

**Results**

**The MOI system.** The MOI system is based on an optical microscope system enhanced by self-customized software (**Fig. 1**). The optical microscope system enables the collection of the bright-field photograph of 2D structures at different magnification. The self-customized software further realizes the intelligent identification of as-collected photograph according to a pre-established database and model. The intelligent identification can be sorted in two steps (**Fig. 1**), *that is*, a training process and a test process. The purpose of the training process is to establish a database and the corresponding SVM model containing the "fingerprint" or characteristic information of RGB channel intensities in the optical photograph of 2D materials with different thicknesses. During the training process, RGB data in optical microscopy photographs of graphene or $MoS_2$ samples with different light intensities ("Training Set" in **Fig. 1**), is manually linked to graphene or $MoS_2$ with different numbers of layer and then classified into different categories by a SVM algorithm, following the judgement of AFM and Raman spectroscopy. The database and SVM model with different categories of RGB channel intensity linked to the sample thickness ("Training Result" in **Fig. 1**) thus makes the following test process possible. During the test process, RGB information in the photographs of graphene or $MoS_2$ ("Test Set" in **Fig. 1**) is collected and classified by the software into specific category. These classifications result in a false-colour image ("Test Result" in **Fig. 1**), which indicates the distribution of substrate, 2D material (with different number of layer) and even impurities. Such a self-customized system inheriting the in-situ and wide-field characterization feature of optical microscopy thus enables accurate, intelligent and large-area identification of 2D structures.

The accurate and intelligent identification relies on the optical contrast characteristics of 2D materials[22-25], as well as the efficient processing and recognition by the self-customized software[32-35]. There are some key features of the MOI system, which can greatly improve the performance of identification. The first one is the pretreatment of the photographs before they are analysed. The pretreatment includes denoising by mean filtering and median filtering, together with colour calibration by linear scaling of G and B channels according to the R channel of the substrate. The colour calibration eliminates possible influence induced by the instability of the optical microscope system. The second issue is about the SVM algorithm. For a small set of training samples, the SVM algorithm is an efficient supervised learning models for data classification[36]. In the MOI system, the SVM classifiers represent the training data of RGB information in a three-dimensional space (see Supplementary Fig. S1), and decide the maximum-margin planes (*i.e.*, boundaries of different categories). After mapping the testing sample into the same space, the maximum-margin plane is evaluated to perform the classification into a specific category (*e.g.*, single-layered graphene). The third one is to perform the identification by multi-channel information of RGB data rather than judge the number of layer by only one channel. Different to conventional identifications based on the contrast different of specific channel between the substrate and the 2D material[25,37], which can suffer from the spatial-inhomogeneity of the light intensity, our multi-channel identification realizes intelligent identification of 2D materials and their numbers of layer, as well as the substrate. Such a multi-channel identification method relies on the optical characteristics of 2D materials in RGB space (see Supplementary Fig. S2 for the results of graphene and $MoS_2$), which implies the possibility to simultaneously identify impurities, 2D materials and even 2D heterostructures without the information of light intensity distribution in the optical microscopy photograph.

**Identification of graphene.** Based on the MOI system, accurate and intelligent identification of graphene sample is demonstrated (**Fig. 2**), which can benefit abundant research and applications of graphene[2-4]. In the training process, optical microscopy photographs of graphene samples (see **Fig. 2a & 2c** for two typical samples) with deterministic judgment of the thickness by the AFM results (**Fig. 2b & 2d**) are processed as a database with different categories of RGB channel intensity linked to the sample thickness (*i.e.*, different numbers of layer) and further analysed by the SVM algorithm to establish a training model. The training result (*i.e.*, the SVM model) contains as-classified RGB information of graphene and the substrate (**Fig. 2e**), which enable the following intelligent identifications of graphene thickness. For a mixed-layer graphene sample (see **Fig. 2f** for the optical microscopy photograph), the MOI system automatically refers the photograph to the training result by analysing the RGB information. As present in **Fig. 2h**, accurate assignments of the number of graphene layer with a pixel-to-pixel accuracy of 96.78% are realized in agreement with AFM results (the inset of **Fig. 2f** and **Fig. 2g**), where regions of different layers are coloured respectively.

**Identification of MoS$_2$.** In general, the MOI method also works for other 2D materials which also have characteristics in RGB space, for example, MoS$_2$ and other transition metal dichalcogenides[22,37]. Accurate and intelligent identification of MoS$_2$ sample is demonstrated in **Fig. 3**. Following a similar training process of graphene, RGB information in optical microscopy photographs of MoS$_2$ samples (**Fig. 3a & 3c**) is collected and analysed by the SVM algorithm according to the thickness judgement by AFM measurements (**Fig. 3b & 3d**). As a result, a training result containing the characteristic RGB information of MoS$_2$ (**Fig. 3e**) can be achieved. As shown in the false-colour image in Fig. 3h, intelligent identification of MoS$_2$ sample can thus be realized automatically with a pixel-to-pixel accuracy of 94.26% based on its optical microscopy photograph (**Fig. 3f**). Besides, the intelligent identification result is also sensitive to impurities or contaminations (black regions in **Fig. 3h**) which can severely affect the intrinsic property of 2D materials[15,38-40]. For example, adhesive residues appearing as light green regions on the substrate and encircling the MoS$_2$ flakes in the optical microscopy photograph (**Fig. 3f**) can be successfully recognized.

**Identification of 2D heterostructure.** With great flexibility in the junction fabrication and property engineering, heterostructures of 2D materials enable the exploration of emerging 2D physics and novel device applications[5-8]. Our intelligent identification method of 2D materials by their "fingerprints" in RGB channels, as a result, may further realize the identification of 2D heterostructures and boost the development of 2D science. **Fig. 4a** shows a 2D heterostructure with a vertical heterojunction of bilayer graphene and single-layered $MoS_2$, which is fabricated following a reported transfer method (see ref.41 for details of the transfer method). Based on the training results of graphene and $MoS_2$ samples (see **Fig. 2e**, **Fig. 3e** and Supplementary Fig. S2), intelligent identification of a graphene-$MoS_2$ heterostructure has been demonstrated (**Fig. 4c**). Regions of substrate, graphene, $MoS_2$, heterojunction, as well as the resist residues from the transfer process can be automatically recognized with a pixel-to-pixel accuracy of 90.16%.

Detailed analyses of the RGB information in different regions (**Fig. 4d & 4e**) indicate that $MoS_2$ dominates the optical contrast of the heterojunction, which can hamper accurate identifications of heterostructures by optical methods. According to the theoretical calculation results (see Supplementary Fig. S3 and refs.25,42-44), further optimization of the oxidation layer thickness and the light wavelength can be adopted to improve the identification performance and even realize the identification of stacking order in the vertical heterojunction. Besides, further comparison of the RGB information before and after the transfer process (**Fig. 4e**) implies the feasibility of this intelligent identification method to evaluate the performance especially the resist residues of 2D material transfer method.

In summary, we report a successful application of machine-learning strategy in the optical identification of 2D structures, including graphene, molybdenum disulphide (MoS$_2$) and their heterostructures. The machine-learning optical identification method (MOI method) relies on trainable and automatic analyses of the red, green and blue (RGB) information in the optical photograph of 2D structures by a support vector machine (SVM) algorithm. By endowing optical microscopy with intelligent insight into the characteristic colour information of 2D materials and 2D heterostructures, the MOI method can realize accurate, intelligent and large-area characterizations of 2D structures, including the thickness, the existence of resist residues, and even the stacking order in heterostructures. With applicability to wafer-scale 2D materials and 2D heterostructures, this intelligent identification method can certainly promote the fundamental research and commercial-level application of 2D structures.

## Methods

**Sample preparation and characterization.** The graphene and MoS$_2$ samples are fabricated by mechanical exfoliation on silicon wafers with 300 nm oxidation layers.[45] The heterostructure samples are prepared by a transfer and stacking method.[41]

Characterizations of the samples are performed by atomic force microscope (Bruker, MultiMode 8), Raman spectroscope (Horiba, Jobin-Yvon LabRAM HR800) and optical microscope (Leica, DM2700 M). Self-customized software for the pretreatment, training and testing processes is completed with MATLAB.

**Training process**. All 2D samples have been photographed in different positions and light intensities to constitute the training set. Each photograph is then divided into sub-regions by the contour of the light intensity which is obtained from a photograph of the substrate (see Supplementary Fig. S4) for the purpose of collecting more RGB information of the sample at different light intensity. Median and mean filters are first adopted in each picture for noise reduction followed by a colour calibration treatment, *that is*, scaling the G, B channel of all pixels in the whole photograph linearly base on a block of manually selected substrate as described above (see Supplementary Fig. S5). Then, using a manually drawn mask for each photograph with the category for corresponding pixels (*e.g.*, single layered graphene region), the mean value of each channel with the same category in one sub-region is calculated as the training dataset. Finally, a three dimensional one-*vs*-one linear SVM can be trained to establish the model containing the characteristic RGB information (*i.e.*, the "Training Result") using the training dataset.

**Test process**. For the test process, denoising and colour calibration treatments are also applied. The category of each pixel can then be automatically identified referring to the training result base on their characteristic RGB information.

**REFERENCES**

1. Das Sarma, S., Adam, S., Hwang, E. H. & Rossi, E., Electronic transport in two-dimensional graphene. *Rev. Mod. Phys.* **83**, 407-470 (2011).

2. Geim, A. K., Graphene: status and prospects. *Science* **324**, 1530-1534 (2009).

3. Geim, A. K. & Novoselov, K. S., The rise of graphene. *Nat. Mater.* **6**, 183-191 (2007).

4. Lin, X. *et al.*, Gate-driven pure spin current in graphene. *Phys. Rev. Appl.* **8**, 034006 (2017).

5. Novoselov, K. S., Mishchenko, A., Carvalho, A. & Castro Neto, A. H., 2D materials and van der Waals heterostructures. *Science* **353**, aac9439 (2016).

6. Xia, F., Wang, H., Xiao, D., Dubey, M. & Ramasubramaniam, A., Two-dimensional material nanophotonics. *Nat. Photonics* **8**, 899-907 (2014).

7. Geim, A. K. & Grigorieva, I. V., Van der Waals heterostructures. *Nature* **499**, 419-425 (2013).

8. Zhang, Z. *et al.*, Robust epitaxial growth of two-dimensional heterostructures, multiheterostructures, and superlattices. *Science* **357**, 788-792 (2017).

9. Wu, T. *et al.*, Fast growth of inch-sized single-crystalline graphene from a controlled single nucleus on Cu–Ni alloys. *Nat. Mater.* **15**, 43-47 (2015).

10. Xu, X. *et al.*, Ultrafast epitaxial growth of metre-sized single-crystal graphene on industrial Cu foil. *Science Bulletin* **62**, 1074-1080 (2017).

11. Lee, J. H. *et al.*, Wafer-scale growth of single-crystal monolayer graphene on reusable hydrogen-terminated germanium. *Science* **344**, 286-289 (2014).

12. Huang, P. Y. *et al.*, Grains and grain boundaries in single-layer graphene atomic patchwork quilts. *Nature* **469**, 389-392 (2011).

13. Meyer, J. C. *et al.*, The structure of suspended graphene sheets. *Nature* **446**, 60-63 (2007).

14. Zhao, W. *et al.*, Low-energy transmission electron diffraction and imaging of large-area graphene. *Science Advances* **3**, e1603231 (2017).

15. Lin, X. *et al.*, Development of an ultra-thin film comprised of a graphene membrane and carbon nanotube vein support. *Nat. Commun.* **4**, 2920 (2013).

16. Zhang, Y. *et al.*, Direct observation of a widely tunable bandgap in bilayer graphene. *Nature* **459**, 820-823 (2009).

17. Wastl, D. S., Weymouth, A. J. & Giessibl, F. J., Atomically resolved graphitic surfaces in air by atomic force microscopy. *ACS Nano* **8**, 5233-5239 (2014).

18. Tu, Q. *et al.*, Quantitative subsurface atomic structure fingerprint for 2D materials and heterostructures by first-principles-calibrated contact-resonance atomic force microscopy. *ACS Nano* **10**, 6491-6500 (2016).

19. Ferrari, A. C. & Basko, D. M., Raman spectroscopy as a versatile tool for studying the properties of graphene. *Nat. Nanotechnol.* **8**, 235-246 (2013).

20. Malard, L. M., Pimenta, M. A., Dresselhaus, G. & Dresselhaus, M. S., Raman spectroscopy in graphene. *Phys. Rep.* **473**, 51-87 (2009).

21. Duong, D. L. *et al.*, Probing graphene grain boundaries with optical microscopy. *Nature* **490**, 235-239 (2012).

22. Li, H. *et al.*, Rapid and reliable thickness identification of two-dimensional nanosheets using optical microscopy. *ACS Nano* **7**, 10344-10353 (2013).

23. Li, W., Moon, S., Wojcik, M. & Xu, K., Direct optical visualization of graphene and its nanoscale defects on transparent substrates. *Nano Lett.* **16**, 5027-5031 (2016).

24. Blake, P. *et al.*, Making graphene visible. *Appl. Phys. Lett.* **91**, 063124 (2007).

25. Ni, Z. H. *et al.*, Graphene thickness determination using reflection and contrast spectroscopy. *Nano Lett.* **7**, 2758-2763 (2007).

26. Li, X. S. *et al.*, Large-area synthesis of high-quality and uniform graphene films on copper foils. *Science* **324**, 1312-1314 (2009).

27. Reina, A. *et al.*, Large area, few-layer graphene films on arbitrary substrates by chemical vapor deposition. *Nano Lett.* **9**, 30-35 (2009).

28. Lee, Y. *et al.*, Synthesis of wafer-scale uniform molybdenum disulfide films with control over the layer number using a gas phase sulfur precursor. *Nanoscale* **6**, 2821 (2014).

29. Zhao, M. *et al.*, Large-scale chemical assembly of atomically thin transistors and circuits. *Nat. Nanotechnol.* **11**, 954-959 (2016).

30. Alexeev, E. M. *et al.*, Imaging of interlayer coupling in van der Waals heterostructures using a bright-field optical microscope. *Nano Lett.* **17**, 5342-5349 (2017).

31. Tan, Y. *et al.*, Tuning of interlayer coupling in large-area graphene/WSe$_2$ van der Waals heterostructure via ion irradiation: optical evidences and photonic applications. *ACS Photonics* **4**, 1531-1538 (2017).

32. Simonyan, K. & Zisserman, A., Very deep convolutional networks for large-scale image recognition. **arXiv:1409.1556**, (2014).

33. Nolen, C. M., Denina, G., Teweldebrhan, D., Bhanu, B. & Balandin, A. A., High-throughput large-area automated identification and quality control of graphene and few-layer graphene films. *ACS Nano* **5**, 914-922 (2011).

34. Poplin, R., Varadarajan, A. V., Blumer, K., Liu, Y. & Mcconnell, M. V., Predicting cardiovascular risk factors from retinal fundus photographs using deep learning. **arXiv:1708.09843**, (2017).

35. Maxmen, A., Deep learning sharpens views of cells and genes. *Nature* **553**, 9-10 (2018).

36. Steinwart, I. & Christmann, A., *Support vector machines*. (Springer-Verlag, New York, NY, 2008).

37. Castellanos-Gomez, A., Agraït, N. & Rubio-Bollinger, G., Optical identification of atomically thin dichalcogenide crystals. *Appl. Phys. Lett.* **96**, 213116 (2010).

38. Chen, J. H. *et al.*, Charged-impurity scattering in graphene. *Nat. Phys.* **4**, 377-381 (2008).

39. Zhu, F. *et al.*, Heating graphene to incandescence and the measurement of its work function by thermionic emission method. *Nano Res.* **7**, 553-560 (2014).

40. Castro Neto, A. H., Guinea, F., Peres, N., Novoselov, K. S. & Geim, A. K., The electronic properties of graphene. *Rev. Mod. Phys.* **81**, 109-162 (2009).

41. Lu, Z. *et al.*, Universal transfer and stacking of chemical vapor deposition grown two-dimensional atomic layers with water-soluble polymer mediator. *ACS Nano* **10**, 5237-5242 (2016).

42. Knittl, Z., *Optics of thin films: an optical multilayer theory*. (Wiley, London, 1976).

43. Zhang, H. *et al.*, Measuring the refractive index of highly crystalline monolayer MoS2 with high confidence. *Sci. Rep.* **5**, 8440 (2015).

44. Palik, E. D., *Handbook of optical constants of solids*. (Elsevier, 1997).

45. Novoselov, K. S. *et al.*, Electric field effect in atomically thin carbon films. *Science* **306**, 666-669 (2004).

**ABBREVIATIONS**

Two-dimensional materials, 2D materials

Molybdenum disulphide, $MoS_2$

Machine-learning optical identification method, MOI method

Red, green and blue information, RGB information

Transmission electron microscopy, TEM

Scanning tunnelling microscopy, STM

Atomic force microscopy, AFM

Photoluminescence, PL

Support vector machine algorithm, SVM algorithm

## ACKNOWLEDGMENT

This work was supported by the National Natural Science Foundation of China (Nos. 51602013, 61602022 and 61627813), the International Collaboration 111 Project (No. B16001), Beijing Natural Science Foundation (No. 4162039) and funding support from Beijing Advanced Innovation Center for Big Data and Brain Computing (BDBC). The authors thank Ms. X.Y. Wang and Prof. Y. Lu for valuable discussions.

## AUTHOR CONTRIBUTIONS

X.Y.L., Z.Z.S., W.Z.F. and J.L.Y. contributed equally to this work. X.Y.L. and W.S.Z. coordinate the project. X.Y.L. proposed and designed the research. Z.Z.S., X.Y.L., X.H.W., S.D.G., Y.C., J.Z., P.L. and K.L.J. performed the sample preparation, characterization and optical photograph collection. W.Z.F, Z.Z.S., J.L.Y. and X.Y.L. performed the image processing and identification. Z.Z.S., X.H.W. and X.Y.L. performed the theoretical calculations. X.Y.L., Z.Z.S., W.Z.F. and J.L.Y. wrote the paper. All the authors participated in discussions of the research.

## Additional information

**Supplementary Information** accompanies this paper at

http://www.nature.com/naturecommunications

## COMPETING FINANCIAL INTERESTS STATEMENT

The authors declare no competing financial interests.

**FIGURES AND LEGENDS**

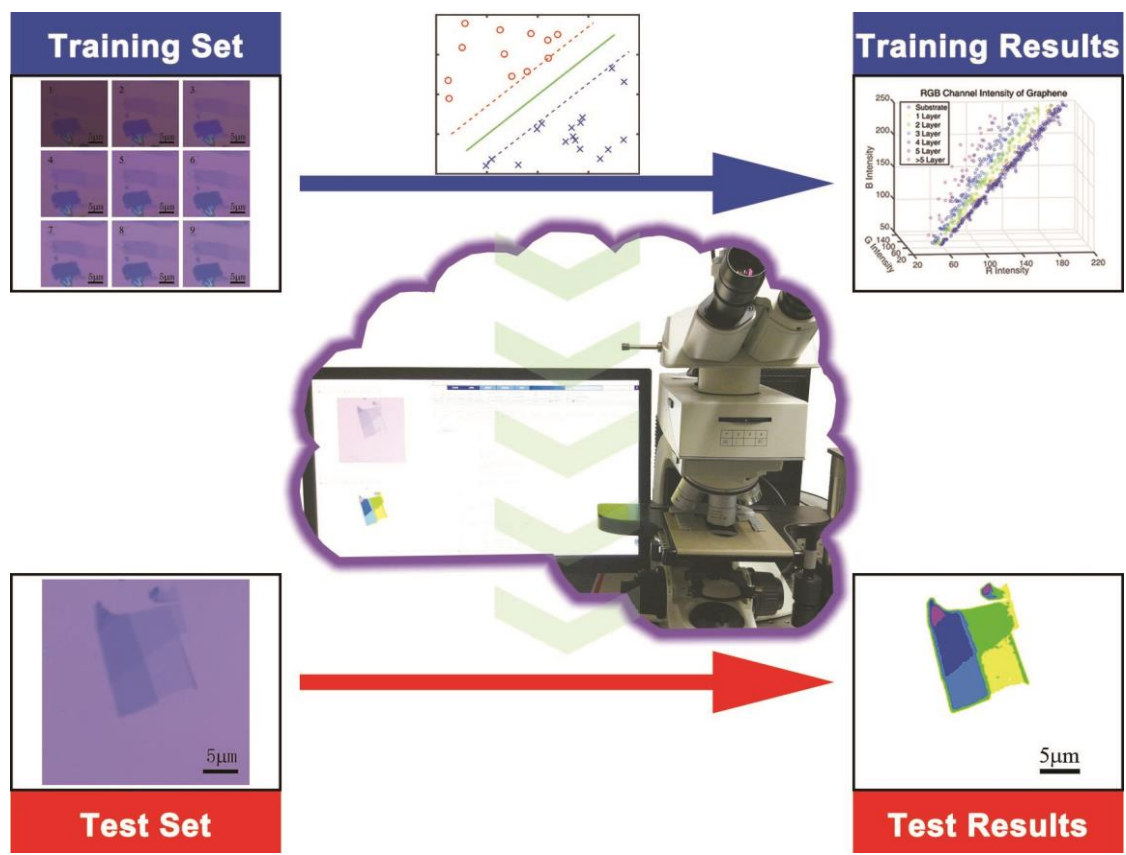

**Figure 1 | The machine-learning optical identification system (MOI system).** Schematic illustration and photograph of the MOI system. The training set contains optical microscopy photographs of graphene or MoS$_2$ samples with different light intensities. Following the judgement of AFM and Raman spectroscopy, RGB database and SVM model of graphene or MoS$_2$ samples (denoted as "Training Results") is established after SVM analyses of the RGB data collected from the training set. Referring to the "Training Results", graphene, MoS$_2$ or heterostructures of these two materials can be identified according to their optical microscopy photograph (denoted as "Test Results"). The MOI system includes an optical microscope enhanced by self-customized software.

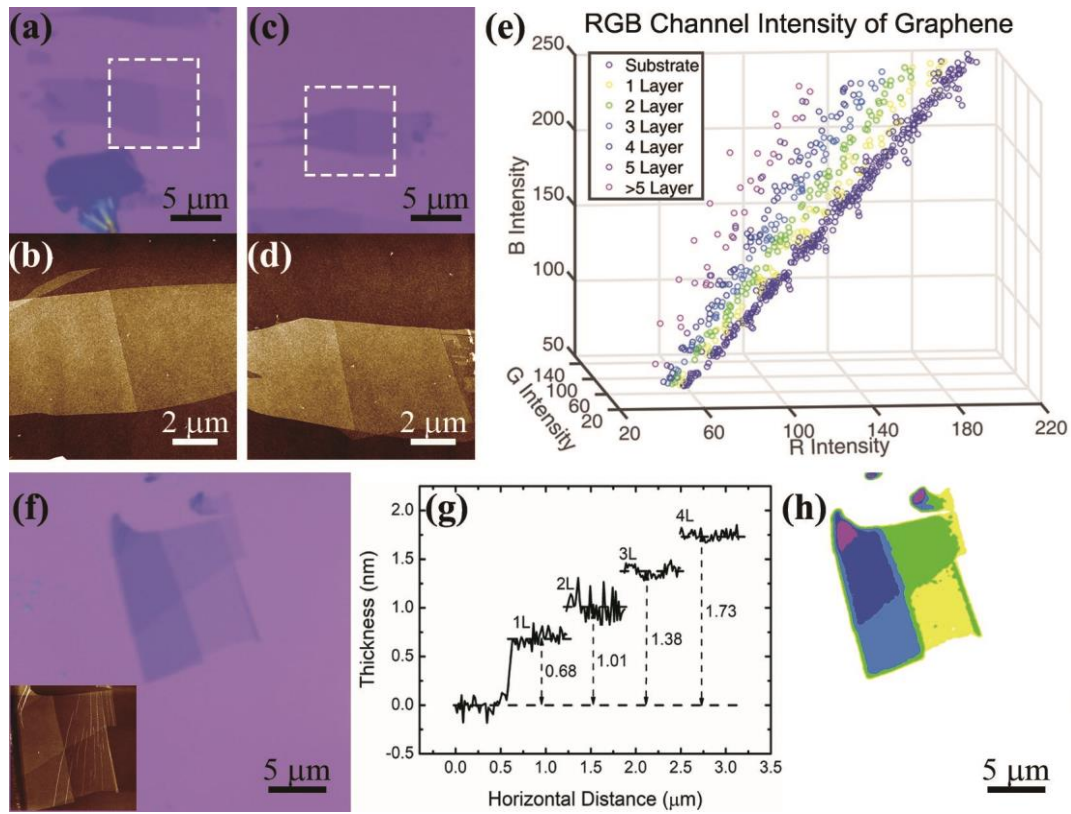

**Figure 2 | Machine-learning optical identification of graphene sample.** (**a-d**) Typical optical microscopy photographs (**a & c**) included in the training set of graphene, where corresponding AFM images (**b & d**) are also present. (**e**) Training result of graphene samples containing as-classified RGB information. Note that, the SVM model is not shown. (**f**) Optical microscopy photograph of a mixed-layer graphene sample for test purpose. Inset shows the corresponding AFM image. (**g**) Thickness information of different layer graphene in (f) by AFM analysis. (**h**) Test result of the sample in (f) according to the database in (e), where regions of different layers are coloured in accordance with (e).

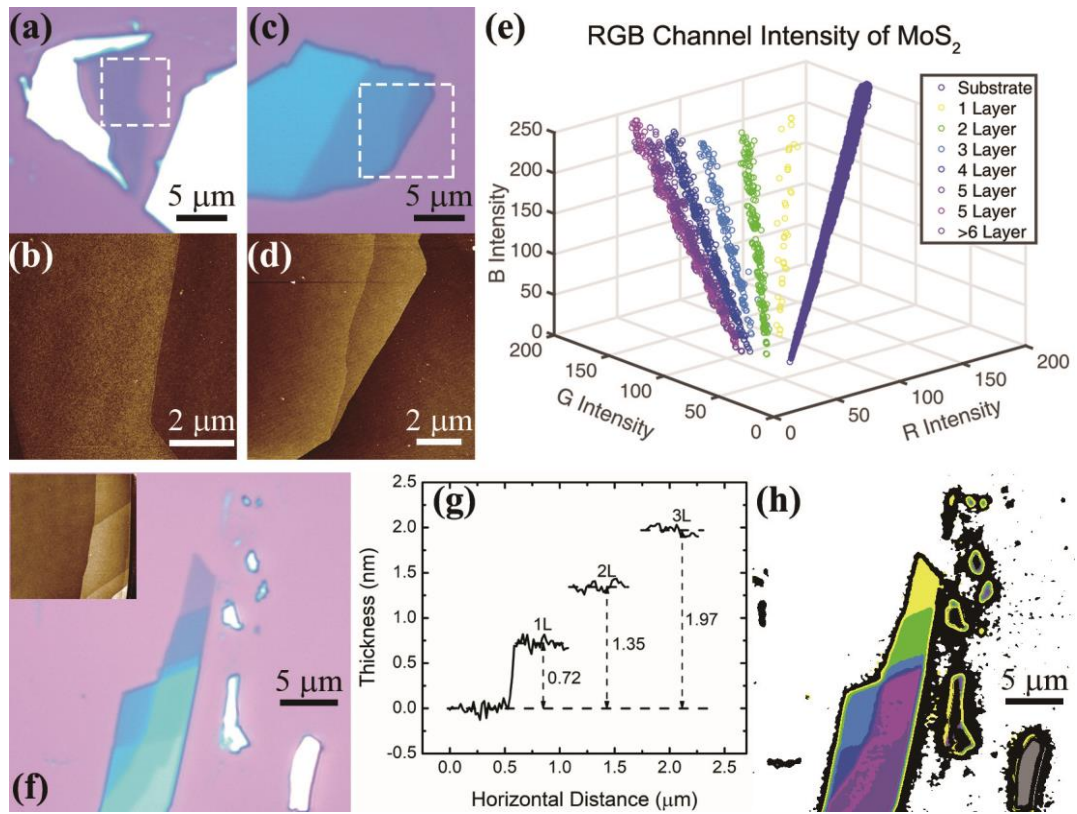

**Figure 3 | Machine-learning optical identification of MoS$_2$ sample.** (**a-d**) Typical optical microscopy photographs (**a & c**) included in the training set of MoS$_2$, where corresponding AFM images (**b & d**) are also present. (**e**) Training result of MoS$_2$ samples containing as-classified RGB information. Note that, the SVM model is not shown. (**f**) Optical microscopy photograph of a mixed-layer sample for test purpose. Inset shows the corresponding AFM image. (**g**) Thickness information of different layer MoS$_2$ in (f) by AFM analysis. (**h**) Test result of the sample in (f) according to the database in (e), where regions of different layers are coloured in accordance with (e). As-identified regions of adhesive residues are blacked. Overexposure regions are greyed.

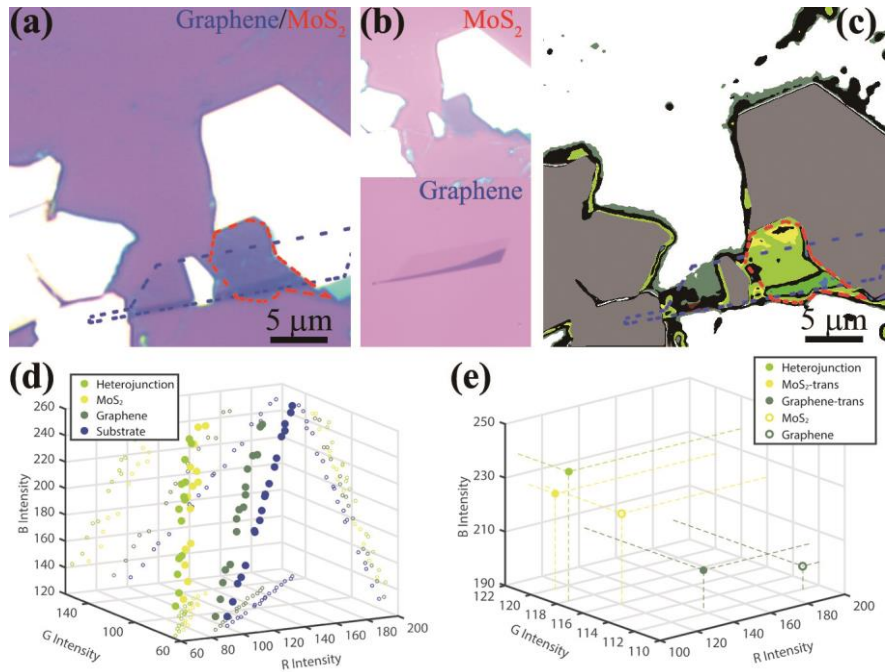

**Figure 4 | Machine-learning optical identification of 2D heterostructure sample.**
(**a**) A 2D heterostructure with a vertical heterojunction of bilayer graphene and single-layered $MoS_2$, where the graphene and $MoS_2$ are marked respectively. (**b**) $MoS_2$ sample and graphene sample used to fabricate the heterostructure. (**c**) Test result of the heterostructure according to the graphene and MoS2 training results, where the graphene and $MoS_2$ are marked respectively. As-identified regions of adhesive residues are blacked. (**d**) RGB information of heterojunction, graphene and $MoS_2$ with different light intensity, where projections of three-dimensional (R,G,B) data onto two-dimensional plane (*e.g.*, RG) are also plotted as circles. (**e**) Comparison of the RGB data of heterojunction, graphene and $MoS_2$ with the same light intensity, together with RGB data of graphene and $MoS_2$ before and after the transfer process.